# Correction of high-order phase variation effects in dynamic field monitoring

Paul I. Dubovan[1,2], Kyle M. Gilbert[1,2], Corey A. Baron[1,2]

[1] Department of Medical Biophysics, Western University, London, Ontario, Canada

[2] Centre for Functional and Metabolic Mapping, Western University, London, Ontario, Canada

**Conflict of Interest**
The authors declare that there is no conflict of interest.

\* Corresponding Author
Name: Paul I. Dubovan
Address: Centre for Functional and Metabolic Mapping
Western University
1151 Richmond St.
London, Ontario, Canada
N6A 3K7
Tel: 519-992-7794
E-mail: pdubovan@uwo.ca

**Funding Information**

NSERC Discovery Grant, Grant Number: RGPIN-2018-05448; Canada Research Chairs, Number: 950-231993; Canada First Research Excellence Fund to BrainsCAN; Canada Foundation for Innovation 37907; Ontario Research Fund 37907; NSERC PGS D program

**Word Count**

4364

Submitted to Magnetic Resonance in Medicine

## Abstract

**Purpose:** Field monitoring measures field perturbations, which can be accounted for during image reconstructions. In certain field monitoring environments, significant phase deviations can arise far from isocenter due to the finite extent of the gradient and/or main magnet. This can degrade the accuracy of field dynamics when field probes are placed near or outside the diameter spherical volume of the gradient coils and/or main magnet, leading to corrupted image quality. The objective of this work was to develop a correction algorithm that reduces errors from highly nonlinear phase variations at distant field probes in field dynamic fits.

**Methods:** The algorithm is split into three components. Component one fits phase coefficients one spatial order at a time, while the second implements a weighted least squares solution based on probe distance. After initial fitting, component three calculates phase residuals and removes the phase for distant probes before re-fitting. Two healthy volunteers were scanned on a head-only 7T MRI using diffusion-weighted single-shot spiral and EPI sequences and field monitoring was performed. Images were reconstructed with and without phase coefficient correction and compared qualitatively.

**Results:** The algorithm was able to correct corrupted field dynamics, resulting in image quality improvements. Significant artefact reduction was observed when correcting higher order fits, especially for diffusion weighted images. Stepwise fitting provided the most correction benefit, which was marginally improved when adding weighted least squares and phase residual corrections.

**Conclusion:** The proposed algorithm can mitigate effects of phase errors in field monitoring, providing improved reliability of field dynamic characterization.

## 1 | Introduction

MRI has become the imaging modality of choice for the diagnosis of many pathologies largely due to its non-invasiveness, excellent soft tissue contrast, and ability to produce high quality anatomical and functional images. However, reconstructed images can often suffer from varying degrees of artefacts and imperfections resulting from magnetic field perturbations that manifest during scan acquisitions. These perturbations can originate from acquisition and system-related effects, such as eddy currents (1,2), heating (3), and mechanical vibration (4), as well as from patient-induced effects, predominantly from motion and breathing (5–7). The ability to reduce artefacts depends on how well field perturbations can be corrected or accounted for, which is particularly important to improve diagnostic efficacy using diffusion imaging, which generates strong eddy currents.

Field monitoring (FM) using field probes (8–12) has demonstrated great utility for measuring dynamic field perturbations that can be incorporated in image reconstructions to correct artefacts. One particular commercially available FM system is composed of sixteen NMR field probes and can measure perturbations up to third order in space (13,14), which is advantageous in correcting higher order perturbations that commonly originate from complex field evolution such as aforementioned eddy currents (15) and patient motion (16).

In order to accurately characterize patient-induced and time-dependent effects, monitoring must be performed concurrently with patient scanning, as opposed to sequential monitoring, which characterizes field dynamics in the empty scanner and uses the measurements to correct images from a subsequent identical acquisition (17). The ability to characterize field dynamics in real-time can be accomplished by either removable probe inserts (18) or by integrating field probes into a radiofrequency (RF) coil (19,20,21). The latter approach is advantageous as it enables more streamlined setup workflows and designs that jointly consider both probe and RF geometry.

For concurrent field monitoring, field probes are distributed around the anatomy of interest and, ideally, are situated well inside the diameter spherical volume (DSV) where the main magnet and gradient fields vary homogenously and linearly, respectively. As such, field probe phase accrual would follow the expected linear behaviour with distance from isocenter, and phase fitting procedures that calculate phase coefficients can be implemented to a high degree of accuracy. In instances where field probes may reside near or outside the ideal region—such as when employing high-performance gradient coils with limited DSVs, when imaging over the large FOVs required for whole-body imaging, or when peripheral equipment prevents probes from being placed near the subject—accrued phase may deviate from linearity, leading to solid harmonic coefficient fitting errors when determining field dynamics and a reduction in the quality of higher order field correction during image reconstruction.

In this work, we propose a simple three-component correction algorithm that diminishes solid harmonic coefficient fitting errors introduced when probes are outside the DSV of the gradient coil and/or main magnet (22). We evaluate the algorithm by employing our previously developed RF head coil with integrated field probes (23). In this RF coil, probes were located as close to isocenter as possible, pursuant to design conditions; however, the physical constraints of the RF coil required all probes to be placed in the nonlinear region of the head-only gradient coil. We have therefore used this architecture as an exemplar to evaluate the performance of our correction algorithm. We demonstrate the benefits of incorporating higher order coefficients by comparing the quality of concurrently monitored field dynamics and resulting images before and after correction, for first through third spatial order fits. The performance of the algorithm is demonstrated for both single-shot spiral and EPI acquisitions. Additionally, we compare computed field dynamics and reconstructed images informed by different levels of algorithm correction to assess the efficacy of each correction component. Lastly, we assess the impact of the correction algorithm on sequentially monitored field dynamics that were not heavily influenced by higher order phase variations. The correction algorithm has been made publicly available to promote dissemination.

## 1.1 | Theory

As discussed by Barmet et al. (8) and Wilm et al. (13), the spatially and temporally varying magnetic field magnitude which encodes the MR signal during an acquisition can be expressed in terms of the static field and spatially smooth dynamic field components, which can be expanded using relatively few basis functions — most notably using the set of $N_L$ real-valued solid harmonics. The magnetic field evolution $B(r,t)$ for a field probe located at position $r$ and at time $t$ can then be expressed as (8):

$$|B(r,t)| = \sum_{l=0}^{N_L-1} B_{ref}(r) + c_l(t)b_l(r) \qquad (1)$$

where $B_{ref}(r)$ denotes the local magnetic field offset relative to the field the spectrometer is tuned to for each probe, $b_l$ denotes the l-th out of $N_L$ solid harmonic, and $c_l$ is the corresponding field coefficient.

The time integrated phase evolution is then described by:

$$\phi(r,t) = \sum_{l=0}^{N_L-1} \gamma B_{ref}(r)t + k_l(t)b_l(r) \qquad (2)$$

where $\gamma$ is the respective gyromagnetic ratio, and $k_l$ now denotes the phase coefficients. For the first order linear harmonics (e.g., $b_l(r) = x$), $k_l$ become the traditional k-space coefficients.

To estimate $\phi(r,t)$ from field probe raw data, two preprocessing steps are first performed: (1) the phase of the raw data is extracted and unwrapped and (2) $\gamma B_{ref}(r)t$ is subtracted from the phase. $B_{ref}$ for each probe is estimated using a calibration scan with no gradients applied.

For an array of $N_P$ probes, Eq. (2) can be cast into vector-matrix notation (8):

$$\boldsymbol{\phi}_P(t) = \boldsymbol{P}\boldsymbol{k}(t) \qquad (3)$$

with $\boldsymbol{\phi}_P(t) = \big(\phi(r_1,t), \phi(r_2,t), \ldots, \phi(r_{N_P},t)\big)^T$

$$\boldsymbol{k}(t) = \big(k_0(t), k_1(t), \ldots, k_{N_l-1}(t)\big)^T$$

and $\boldsymbol{P} = \dfrac{\gamma_P}{\gamma} \begin{pmatrix} b_0(r_1) & b_1(r_1) & \cdots & b_{N_L-1}(r_1) \\ \vdots & \vdots & \vdots & \vdots \\ b_0(r_{N_P}) & b_1(r_{N_P}) & \cdots & b_{N_L-1}(r_{N_P}) \end{pmatrix}$

where the different entries of $\boldsymbol{P}$ represent the different solid harmonic basis functions that are sampled by the probe array.

An approximate solution to Equation (3) can be written as:

$$\boldsymbol{k}(t) = \boldsymbol{P}^+ \boldsymbol{\phi}_P(t) \qquad (4)$$

where $\boldsymbol{P}^+ = (\boldsymbol{P}^T \boldsymbol{P})^{-1} \boldsymbol{P}^T$ denotes the Moore-Penrose pseudoinverse of $\boldsymbol{P}$.

Therefore, to obtain the phase coefficients during a scan acquisition, the phase time-courses of each probe are measured during the scan, and Equation (4) is carried out independently at each point in time using probe positions as determined from a separate calibration scan to populate probing matrix $\boldsymbol{P}$.

While not explicitly included in the above equations, the calculation of solid harmonic coefficients also considers the phase contributions from second order concomitant field terms (24). The expected phase from concomitant fields is computed from the first order k terms, which is then subtracted from $\boldsymbol{\phi}_P(t)$ before re-evaluating Eq. 4. This process is done iteratively to arrive at a final solution for $\boldsymbol{k}(t)$ and the time-varying concomitant fields (14,17). For notational simplicity, henceforth $\boldsymbol{P}$ and $\boldsymbol{k}(t)$ are considered to include the concomitant field basis functions and coefficients, respectively, and Eq. 4 is considered to carry out this iterative procedure.

## 2 | Methods

### 2.1 | Correction Algorithm

The full corrected fitting algorithm consists of three components: stepwise-fitting, weighted least squares (WLS) optimization, and weighted phase residual (WPR) removal. A flow chart of the algorithm is depicted in Figure 1, and a portrayal of the correction steps in the form of a one-dimensional exemplar problem is depicted in Figure 2. Both figures are described in more detail below.

#### 2.1.1 | Stepwise Fitting Correction

Conventionally, the coefficients $\boldsymbol{k}(t)$ are calculated simultaneously via Eq. (4) by incorporating all basis functions in $\boldsymbol{P}$. When the probes are near isocenter, a simultaneous higher order fit is able to accurately represent the gradient field. However, in the presence of substantial higher-order phase variation for distant field probes that may be outside the DSV (likely higher than the 3rd order harmonics that typical fits utilize), a simultaneous higher order fit fails to accurately capture the zeroth and first order behaviours of the gradient field because the very high orders that the model is not equipped to fit are projected to lower orders. To understand this in one dimension ($x$) for a gradient with non-linearity, if many probes sample the phase in the non-linear region, a second order fit up to $x^2$ inherits significant higher order behaviour and does an inadequate job at fitting the true linear behaviour of the gradient field near isocenter (Figure 2). To avoid this underfitting of zeroth and first order terms, we propose solving $\boldsymbol{k}(t)$ one order at a time, which acts as an implicit regularization that strongly favors the zeroth and first order terms. The zeroth and first order terms are computed together in the first step, as is traditionally done for first-order fits, since calculation of the zeroth order term requires knowledge of the probe positions (first-order solid harmonics) to not overfit the zeroth order solution, except in the very ideal case where probes are completely symmetric about isocenter. After calculating $\boldsymbol{k}(t)$ for each order, the residual phase is calculated as per Eq. (5), and is used in the calculation of coefficients for subsequent orders:

$$\boldsymbol{\phi}_{P,res}(t) = \boldsymbol{\phi}_P(t) - \boldsymbol{P}\boldsymbol{k}(t) \quad (5)$$

In the one-dimensional case, performing stepwise correction leads to a fit that much more closely aligns with the ground truth field.

### 2.1.2 | Weighted Least Squares Correction

To further suppress effects from high order phase variation, Eq. (4) was modified to include WLS optimization as follows:

$$\boldsymbol{k}(t) = (\boldsymbol{P}^T\boldsymbol{W}\boldsymbol{P})^{-1}\boldsymbol{P}^T\boldsymbol{W}\boldsymbol{\phi}_P(t) \qquad (6)$$

where $\boldsymbol{W}$ is a diagonal matrix of weights. Here, we choose that the weight for the $j^{th}$ probe depends inversely on Euclidean probe distance from isocenter $r_j$ because, by definition, the higher orders of phase variation that should be suppressed increase in magnitude with distance from isocenter:

$$\boldsymbol{W}_j = 1/\left\|\boldsymbol{r}_j\right\|_2^4 \qquad (7)$$

$$\boldsymbol{W}_j = min_j(\left\|\boldsymbol{r}_j\right\|_2^4)/\left\|\boldsymbol{r}_j\right\|_2^4 \qquad (8)$$

where the numerator normalizes the entries of $\boldsymbol{W}$ to be between 0 and 1. Building off the one-dimensional model, the incorporation of WLS makes the fitting more accurate by enabling probes affected less by nonlinearity to contribute more strongly to the fit.

### 2.1.3 | Weighted Phase Residual Removal

The third correction assumes that any residual phase after fitting using the first two corrections is primarily attributed to errors from unaccounted-for higher orders, especially for the probes far from isocenter. The residual phase is thus removed from the initial probe phases by weighting the residuals that get subtracted using weights of min($\boldsymbol{W}$)/$\boldsymbol{W}$. This causes probes far from isocenter to have a larger correction than probes near isocenter. Heuristically, we saw the best results when using 2-3 iterations for this phase fitting process. As shown in Figure 2, the act of removing residual phase from the probes can be thought of as displacing the sampled phase values such that the phase distribution is of a lower order in space, which in turn leads to a more accurate fitting of the lower order coefficients.

### 2.1.4 | Algorithm Access

The correction algorithm has been implemented in the MatMRI toolbox that is publicly available at https://doi.org/10.5281/zenodo.4495476 and https://gitlab.com/cfmm/matlab/matmri (25), where the FM k-coefficients are computed offline using the raw field probe data files collected with a Skope system during an FM acquisition.

## 2.2 | MR Acquisition

Two healthy volunteers were scanned on a 7T head-only MRI scanner (Siemens Terra Dot Plus) at Western University's Centre for Functional and Metabolic Mapping (80 mT/m gradient strength and 400 T/m/s maximum slew rate). This study was approved by the institutional review board at Western University, and informed consent was obtained before scanning. Diffusion-weighted single-shot spiral acquisitions were performed with parallel imaging acceleration rates of 2,3, and 4. The imaging parameters were as follows: FOV: 192 x 192 mm$^2$, in-plane resolution: 1.5 x 1.5 mm$^2$, slice thickness: 3 mm, number of slices: 10, TE/TR: 33/2,500 ms, flip angle: 70°, b = 0 s/mm$^2$ acquisitions: 1, diffusion directions: 6, b-value: 1000 s/mm$^2$.

A Cartesian EPI diffusion-weighted acquisition was also acquired using the same 6-direction diffusion scheme and the following imaging parameters: acceleration rate: 2, partial Fourier = 6/8, FOV: 192 x 192 mm$^2$, in-plane resolution: 1.5 x 1.5 mm$^2$, slice thickness: 3 mm, number of slices: 10, TE/TR: 45/2,500 ms, flip angle: 70°.

A Cartesian dual-echo gradient-echo acquisition was used to estimate B$_0$ maps for inclusion in a model-based reconstruction to correct for static off-resonance effects. The imaging parameters were as follows: FOV = 240x240 mm$^2$, spatial resolution = 1.5-mm isotropic, TE$_1$/TE$_2$ = 4.08/5.10 ms. From the first echo, sensitivity coil maps were estimated using ESPIRiT (26).

FM was performed concurrently using 16 transmit/receive $^{19}$F commercial field probes (Skope) that are integrated into a 8-channel transmit, 32-channel receive RF head coil (23). Sequential FM of an identical diffusion-weighted single-shot spiral acquisition was also performed prior to integration of the field probes into the RF coil.

## 2.3 | Image Reconstruction and Data Analysis

Image reconstruction was performed in MATLAB using the in-house developed MatMRI toolbox (25) that uses an iterative expanded encoding model-based reconstruction (13). All images were reconstructed using 20 iterations with the conjugate gradient method, and coil compression (27-30) to 20 virtual coils was performed to improve reconstruction speed. Noise correlation between receivers was corrected using prewhitening before any reconstructions (31). Prior to reconstruction, field dynamics were also adjusted to account for vendor-Maxwell corrections and eddy current compensation that are not measured by the field probes, using methods described previously by the authors (23) and reproduced here with notable differences expounded upon. Namely, additional $B_0$ and linear Maxwell terms are produced given the asymmetric gradient coil design (24) and so the MRI scanner corrects for the linear terms automatically by applying dynamic gradients to counteract them and corrects for the $B_0$ terms by applying the appropriate demodulation phase (32). The linear correction was presumed to be measured by the field probes, whereas the DC correction is invisible to the field probe system and was thus reversed during reconstruction using Maxwell term predictions based on the first-order field probe measurements. Similarly, the scanner performs $B_0$ eddy current compensation using similar demodulation which is also invisible to the field probes and needs to be accounted for in measured field dynamics retrospectively. This is done by estimating the zeroth order corrections via offline simulation of the full acquisition (including diffusion gradients) and reversing them during reconstruction in a similar manner to the DC Maxwell terms.

Images were reconstructed using uncorrected and corrected first-order, second-order, and third-order k-coefficient fits. Additionally, image reconstructions using second order fits were performed using no correction algorithm, only stepwise correction, stepwise+WLS, and all three correction techniques (full correction) to evaluate each correction component's contribution to overall correction. Second order was used here instead of 3rd order, because the full correction strategy relies on an overdetermined system of equations (i.e., more probes than spherical harmonic basis functions), which is not satisfied for the 3rd order case where the number of probes is equal to the number of basis functions. Following image reconstruction, fractional anisotropy (FA) maps were computed using the MRtrix3 package (33).

**3 | Results**

Full correction (stepwise + WLS + WPR) of up to third-order k-coefficients acquired from a spiral DWI concurrently monitored trajectory (Fig. 3a) showed significant zeroth to third order trajectory differences compared to the uncorrected k-coefficients (Fig. 3b). Applying the stepwise correction appeared to induce the largest difference in the profiles out of the three correction strategies. When employing a second order fit, the difference between uncorrected and fully corrected terms is smaller yet still apparent. Similarly, stepwise appears to handle most of the profile correction. First order fits showed small yet observable differences in trajectories when adding WLS and WPR correction. Note that stepwise correction is not applicable for first order fits. Comparable results were observed for all basis functions (Fig. S1). Notably, results from the full set of basis functions consistently showed larger uncorrected deviations for terms having z dependency, especially those having $z^2$ and $z^3$ dependency, such as $k_6$ and $k_{12}$, respectively. This observation is expected given that fields rapidly change along the z-direction for head-only MRI and gradient systems when outside the designed linear region. Also, the deviations varied significantly depending on DWI gradient direction due to diffusion gradient eddy currents (Fig. 3c).

Images and FA maps reconstructed with and without k-coefficient correction, for first through third order fits of concurrently monitored FM data, show degrading image quality when employing higher order uncorrected fits (Fig. 4). When correcting the k-coefficients, image quality is improved, particularly for FA maps. This is evidenced by the reduction in blurring and artifactually elevated FA in the grey matter and CSF for higher order fits. FA map zoom-ins show how k-coefficient correction enables the accurate use of second order fits, which leads to improved image quality when compared to images using first order fits. This is illustrated more effectively in the video in Animation S2, which shows apparent DWI blurring reduction and improved FA map integrity when transitioning from first order corrected images to second order corrected images. The zoom-ins and video also enable effective comparisons of images reconstructed using first order fits, which show clear improvements in quality when correcting the k-coefficients. Furthermore, DWI and FA maps informed by corrected third order fits were slightly blurrier and noisier, respectively, than those informed by corrected second order fits. Similar trends were observed for images and FA maps (Fig. 5 & Animation S4), as well as for k-coefficient difference profiles (Animation S4) from the diffusion weighted EPI acquisition.

In Figure 6, images and FA maps reconstructed using second-order concurrent monitored k-coefficient data corrected using only stepwise correction appeared to provide the majority of the correction benefit to the images. Despite this, individually adding WLS and WPR corrections further improved image quality and integrity of FA maps, which is more clearly shown in the zoom-ins, as well as in a video in Animation S3.

Correction of sequentially monitored third-order k-coefficients showed small differences in k-coefficients (Fig. 7a), with the degree of difference being far smaller in magnitude compared to the concurrent case. This is also illustrated via image reconstructions and FA maps in b), which shows negligible differences in image quality when using corrected sequentially monitored k-coefficients, as opposed to the substantial image quality improvements observed for corrected concurrently monitored k-coefficients.

**4 | Discussion**

In this work, we present a solution for correcting computed k-coefficients from field monitoring data for situations in which higher order phase variations significantly affect spherical harmonics fitting, such as the case that arises when field probes are placed outside the homogenous/linear region of the main magnet and/or gradient coil.

When implementing a third-order fit, the conventional inverse problem that solves all orders simultaneously introduces large errors for fitted terms, as evidenced by the large deviations in k-coefficients and significant image artefacts (Figs. 3, 4). The large sensitivity is due to the problem being a balanced system of equations with an equal number of equations (i.e., probes) to unknowns (i.e., spherical harmonic k-coefficients), which causes the probes far from isocenter in regions of high non-linearity to have a large erroneous effect on the solution. Larger errors were exhibited by terms having stronger z-dependency, which is consistent with the rapidly changing fields along the z-direction that are expected for a head-only scanner and gradient coil (for both eddy currents and applied gradients). In other words, spherical harmonic orders greater than 3 are likely required for probes far from isocenter, but there are insufficient probes for this type of model. When the probes are close to isocenter, these problematic high-order terms that scale with the 4th or higher powers of radius become negligible compared to the lower orders. These k-coefficient errors become less apparent as the fit order is reduced, likely due to the system becoming overdetermined, but are still present and possible to mitigate with the proposed fitting algorithm (Fig. 3). While the stepwise correction is the most effective of the three strategies employed, additional image quality improvement is observed with the addition of each of the WLS and WPR corrections (Fig. 6). For optimal performance, we suggest using WPR correction up to fits that are one order lower than the allowable fit, i.e., for a 16-probe system that allows up to third order fits, this correction is expected to perform well up to second order. To be consistent with the correction techniques used, WPR correction was used in correcting third order fits in this work, given there was little difference in image quality when and when not using WPR correction for third order fits. Altogether, the use of these corrections enables the production of images of higher quality when using higher order fits compared to using a first order fit alone (Figs. 4, 5, and better shown in the video in Animation S2). However, in the event that a first order fit is chosen by the user for image reconstruction, WLS and WPR corrections still provide better quality than when not using the correction strategies (Figs. 4, 5, and Animation S4).

Interestingly, images were of slightly lower quality overall when using corrected third-order k-coefficients compared to corrected second-order k-coefficients. This is likely because all three components of the proposed algorithm capitalize on over-determined solutions and a third-order solution has balanced numbers of equations and unknowns for 16 probes. In contrast, the improvements in image quality for second-order corrected versus both first-order corrected and second-order uncorrected (Fig. 4) suggest that the second-order terms are more accurately estimated via the proposed methods. The larger distortions observed for uncorrected DWI compared to non-diffusion weighted images is likely due to diffusion gradient-induced eddy currents, which likely have strong spatial variation along the z-direction given the short spatial extent of the head-only gradient coils and cryostat of the head-only MRI that the eddy currents primarily reside in.

A limitation of this work is the qualitative nature of the assessments. However, a quantitative metric that can capture the subtle changes introduced by the different correction steps and fitting orders is elusive. Nevertheless, the improvements from the algorithm are readily apparent from visual inspection, especially from the FA maps. Notably, FA is highly sensitive to inconsistencies between images with different diffusion gradient encoding directions that result from the errors the proposed algorithm aims to address.

Blurring appears to be present in some images even after the correction, which may be a combination of uncorrected phase errors and $T_2^*$ blurring. However, a modest acceleration rate of 2 was chosen for this work to better demonstrate the effects of the corrections, because errors in k-coefficients are exacerbated by slower traversal through k-space (similar to distortions/artefacts from uncorrected $B_0$ inhomogeneity). When utilizing acceleration rates of 3 or 4, which have been shown to have low g-factors for single shot spiral acquisitions (34), this blurring is greatly reduced (Fig. 8).

When implementing WLS correction, different weighting schemes were tested, including weightings based solely on probe distance in the z-direction, probe signal magnitude, and probe signal decay rate. However, net distance from isocenter was qualitatively most effective in consistently improving image quality. The poor performance for distance along z, even though it is the higher order modes with large z-terms that are most affected by the corrections is due to the fact that probes that are distant in both z and x/y likely reside in regions with greater inhomogeneity than those distant only along z. Interestingly, full removal of the most distant probe(s) consistently reduced image quality, suggesting that even the farthest probes still have useful information to provide during the fits. Nevertheless, there may be more optimal choices for probe weightings.

The negligible difference in image quality when correcting sequential FM data was expected given that all probes were within the vendor-defined DSV of the scanner; accordingly, image reconstructions using sequentially monitored data were already of high quality. These results also suggest that the algorithm is suitable for routine use in sequential or concurrent monitoring environments that do not experience large phase errors, as no reconstruction errors were introduced.

The proposed algorithm requires very few modifications to the original workflow, and negligible computation time is added. The proposed approach is also effective in correcting different trajectories, as shown for a diffusion-weighted EPI trajectory of the same resolution and a comparable readout time (Fig. 5 and Animation S4). The k-coefficient errors present in the EPI acquisition translate to different artefacts than for the spiral acquisition, with bulk shifts in the phase encoding direction, compression, and ghosting being observed. This translates to significantly corrupted FA maps for higher order uncorrected fits, which are corrected using the proposed correction techniques. Notably, similar correction improvements were observed for the second volunteer (Fig. S5).

A fourth correction step was initially investigated. This correction method used vendor-provided gradient non-linearity characterizations (35,36) to adjust the calibrated probe positions, and the x, y, and z k-coefficient basis functions in both the fitting of the probe data and in image reconstruction. However, after evaluating the performance of the correction, the benefit was negligible when combined with the other correction techniques and was not deemed necessary given the additional several minutes of computation time required to correct the probe positions (Fig. S6), and due to the fact that the 3-step algorithm can remain vendor-agnostic and be made entirely open-source as it does not rely on proprietary vendor information. The poor performance of this approach was likely because it does not correct for eddy current modes with rapid spatial variations far from isocenter. Also, the vendor characterization may only be valid within the DSV of the gradient coils. While it may be possible to calibrate the field nonlinearity of the gradients and eddy current modes outside the DSV, it would likely require a lengthy calibration process.

Another potential correction approach could be to introduce a tailored regularization term in Equation 4 that penalizes large contributions of z-dependent spherical harmonic terms. However, this approach would require tuning of the regularization strength for the different basis functions and would require knowledge of the spherical harmonic basis functions that are more prone to error. In contrast, the proposed approach makes no assumptions about the coil/scanner geometry (e.g., perhaps coils with x or y asymmetry may also benefit from it (37)) and does not require tuning of parameters. That said, regularization could also potentially be used in combination with the proposed algorithm, or the proposed algorithm could be used to identify problematic basis functions that could be penalized with regularization, which are promising avenues for future work.

## 5 | Conclusion

A correction algorithm that successfully improves the reliability of phase-corrupted FM data is presented. This encourages the use of FM systems to achieve higher quality images even when probes may lie outside the imaging volume, including complex FM environments such as concurrent FM in head-only scanners and imaging of anatomy distant from isocenter with large FOVs.

**Acknowledgements**

The authors would like to thank the NSERC Discovery Grant [grant number RGPIN-2018-05448], Canada Research Chairs [number 950-231993], Canada First Research Excellence Fund to BrainsCAN, Canada Foundation for Innovation 37907, Ontario Research Fund 37907, and the NSERC PGS D program.

**Data Availability Statement**

The k-coefficient calculation along with the source code used for the expanded encoding model reconstructions are publicly available in the MatMRI toolbox: https://doi.org/10.5281/zenodo.4495476 and https://gitlab.com/cfmm/matlab/matmri (25). A demo calculating phase coefficients from raw probe data acquired from a spiral acquisition with and without the proposed corrections is included.

ORCID

Paul I. Dubovan https://orcid.org/0000-0002-5377-6975
Kyle M. Gilbert https://orcid.org/0000-0003-3026-5686
Corey A. Baron https://orcid.org/0000-0001-7343-5580

TWITTER

Paul I. Dubovan @pdubovan3
Corey A. Baron @cabaron31

MASTADON
Corey A. Baron @cbaron@mrtodon.net

**List of Figures and Captions**

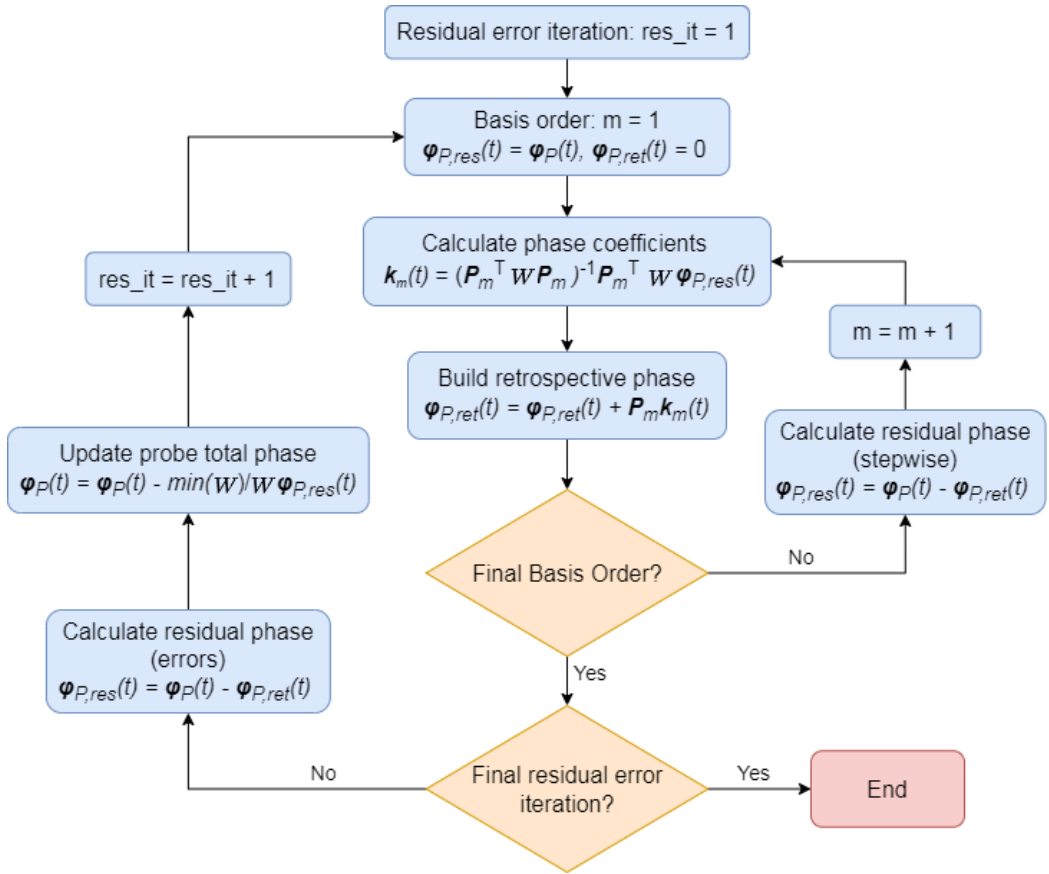

**Figure 1** Flowchart outlining the calculation of phase coefficients from probe phase data while using the proposed correction steps: stepwise correction, weighted least squares (WLS), and weighted phase residuals (WPR). $k_m(t)$ and $P_m$ contain all the coefficients and spherical harmonic bases, respectively, of the $m^{th}$ order. For $m = 1$, the $0^{th}$ and $1^{st}$ order harmonics ($l = \{0,1,2,3\}$) are included in $k_m(t)$ and $P_m$, for $m = 2$ the $2^{nd}$ order harmonics ($l = \{4,5,6,7,8\}$) are included, and so on for higher orders. The weighted vector $W$ is calculated as per Equations (7) and (8), and is reshaped into a diagonal matrix when multiplied by $P_m$. Contribution of accrued phase from concomitant field terms is not explicitly shown in the chart, but is also included during the "calculate phase coefficients" step.

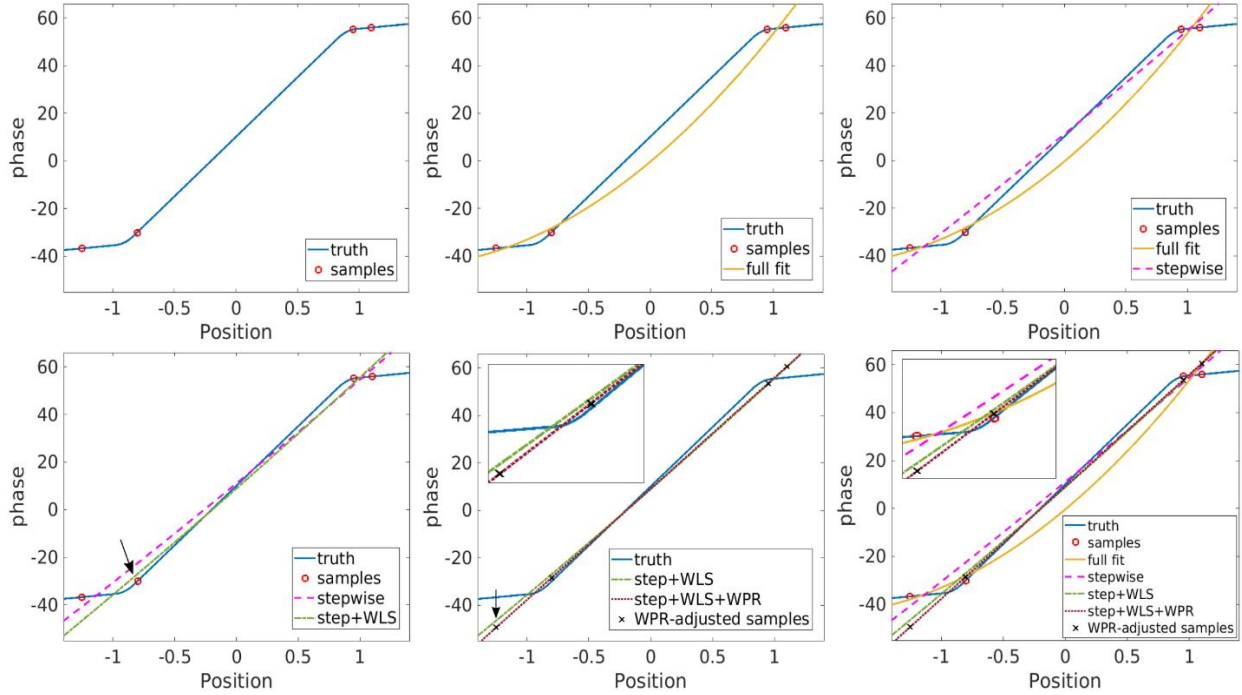

**Figure 2** One-dimensional probe fitting scenario illustrates how fitting a phase profile with probes outside the linear region becomes more accurate when employing the proposed correction strategies. For the arbitrary direction $x$, the basis functions selected to fit the phase were $(1, x, x^2)$. Notably, WPR only significantly adjusts the phase for the probes farthest from isocenter.

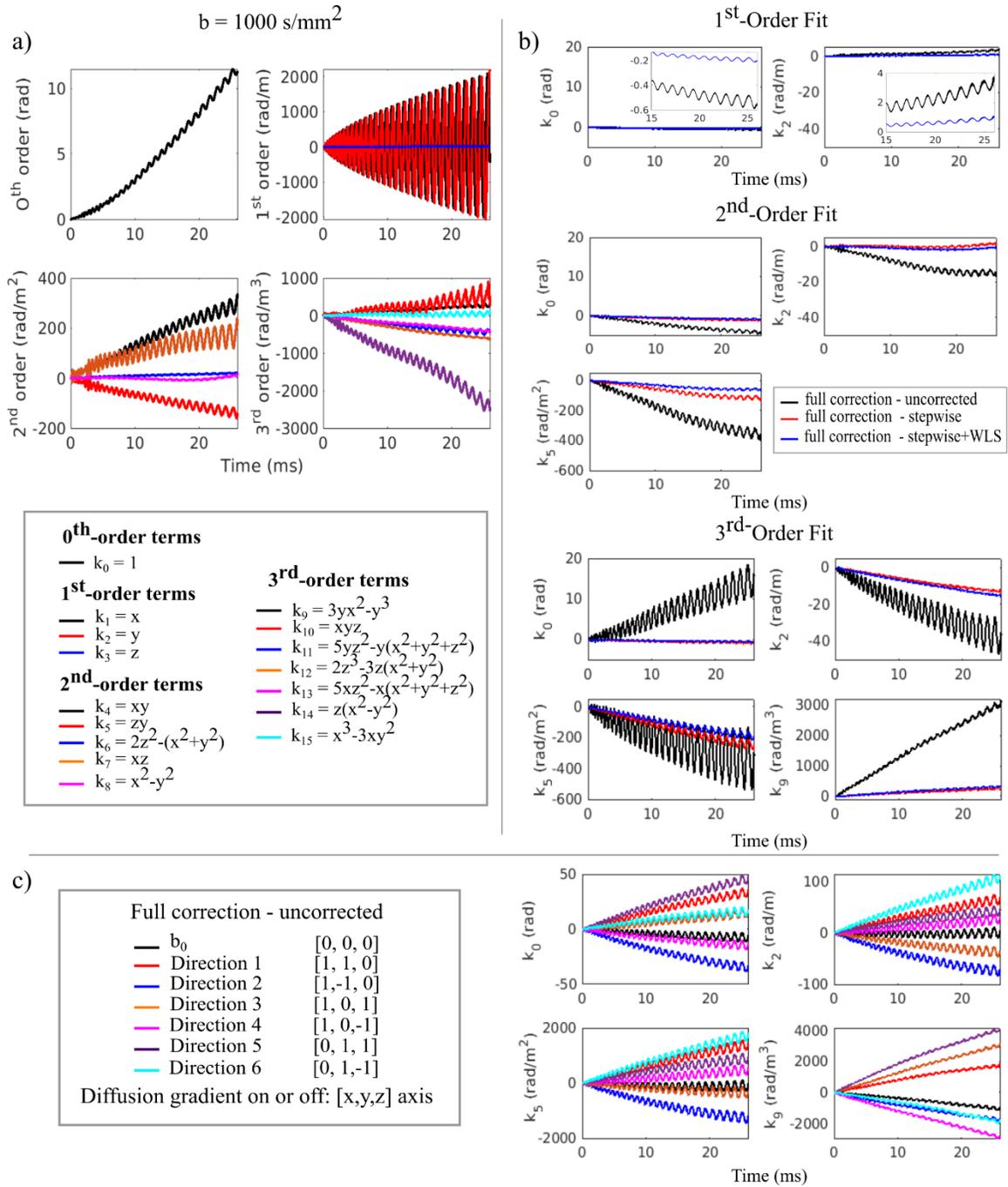

**Figure 3** (a) Fully corrected concurrently monitored k-coefficients generated with a third-order fit from a diffusion weighted single-shot spiral acquisition, with an acceleration rate of 2. (b) Difference between the fully corrected k-coefficients and uncorrected coefficients (black), coefficients corrected using only stepwise correction (red), and coefficients corrected using stepwise + WLS correction (blue), for maximum order of the fits ranging from first through third order. Only black and blue curves displayed for the first order fit since there is no stepwise correction to be performed. Sample basis function displayed for each order. Insets added in the first order plots highlighting the observable differences in the profiles from 15 ms onwards. Full set of basis functions shown in Figure S1 for both non-diffusion and diffusion weighted acquisitions along with uncorrected and fully corrected k-coefficient profiles with no differences taken. (c) Difference profiles of fully corrected and uncorrected k-coefficients for all acquisitions in the diffusion protocol, consisting of 1 b = 0 s/mm$^2$ and 6 diffusion-weighted (b = 1000 s/mm$^2$) acquisitions. Diffusion gradient axes that were on for each acquisition are provided by the respective directions. Large differences observed between the different directions are likely due to diffusion gradient-induced eddy currents.

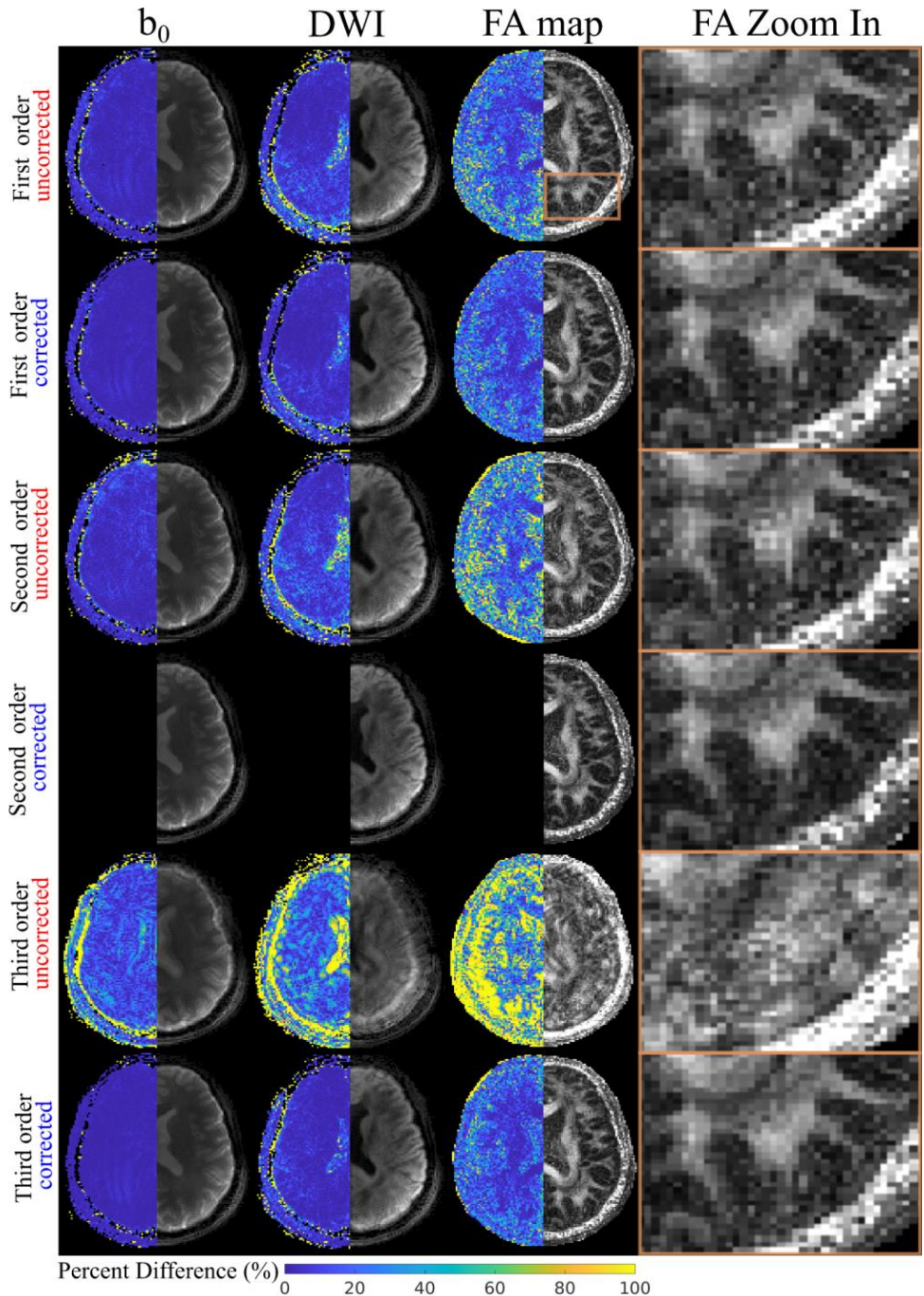

**Figure 4** Non-diffusion weighted images (b = 0 s/mm$^2$, left), single-direction diffusion weighted images (DWI) (b = 1000 s/mm$^2$, middle), and fractional anisotropy (FA) maps (right) reconstructed from single-shot spiral acquisitions (R = 2) using first through third order concurrent monitored field dynamics. Reconstructions incorporated uncorrected k-coefficients and corrected k-coefficients using the proposed stepwise, WLS, and WPR correction techniques. Difference images (left hemisphere) were calculated with respect to images that used fully corrected second order field dynamics, which we qualitatively observed

to have the best image quality. Zoom-ins highlight improved FA map integrity when using corrected first order fits, which are further improved using corrected second order fits. An animation more effectively illustrating these observations is shown in Animation S2.

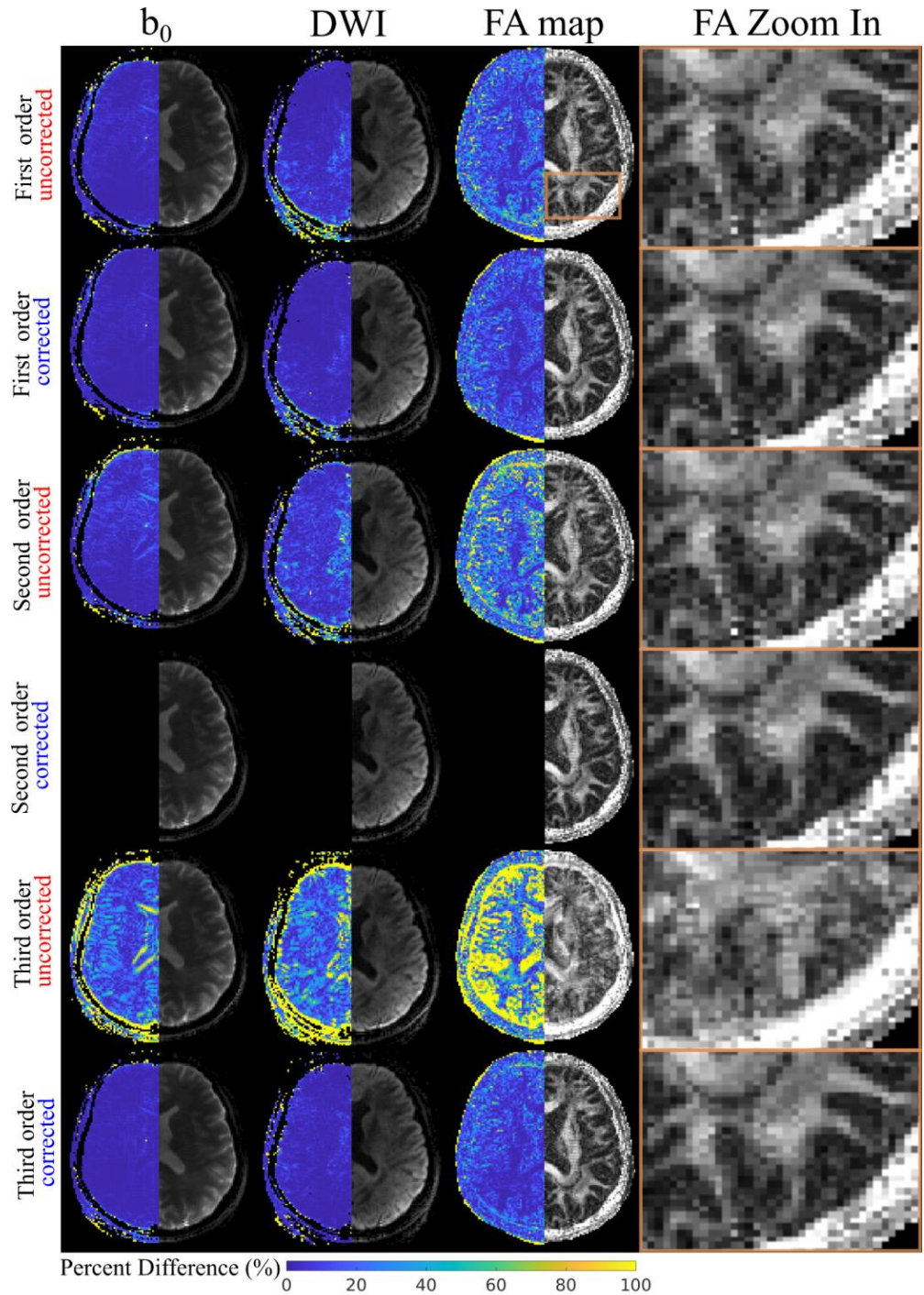

**Figure 5** Non-diffusion weighted images (b = 0 s/mm$^2$, left), single-direction DWI (b = 1000 s/mm$^2$, middle), and FA maps (right) reconstructed from single-shot EPI acquisitions (R = 2) using first through third order concurrent monitored field dynamics. Reconstructions incorporated uncorrected k-coefficients and corrected k-coefficients using the proposed stepwise, WLS, and WPR correction techniques. Difference images (left hemisphere) were calculated with respect to images that used fully corrected second order field dynamics, which we qualitatively observed to have the best image quality. Zoom-ins show improved FA map integrity when using corrected first order fits, which are further improved using corrected second order fits. Also see Animation S4 for an animation.

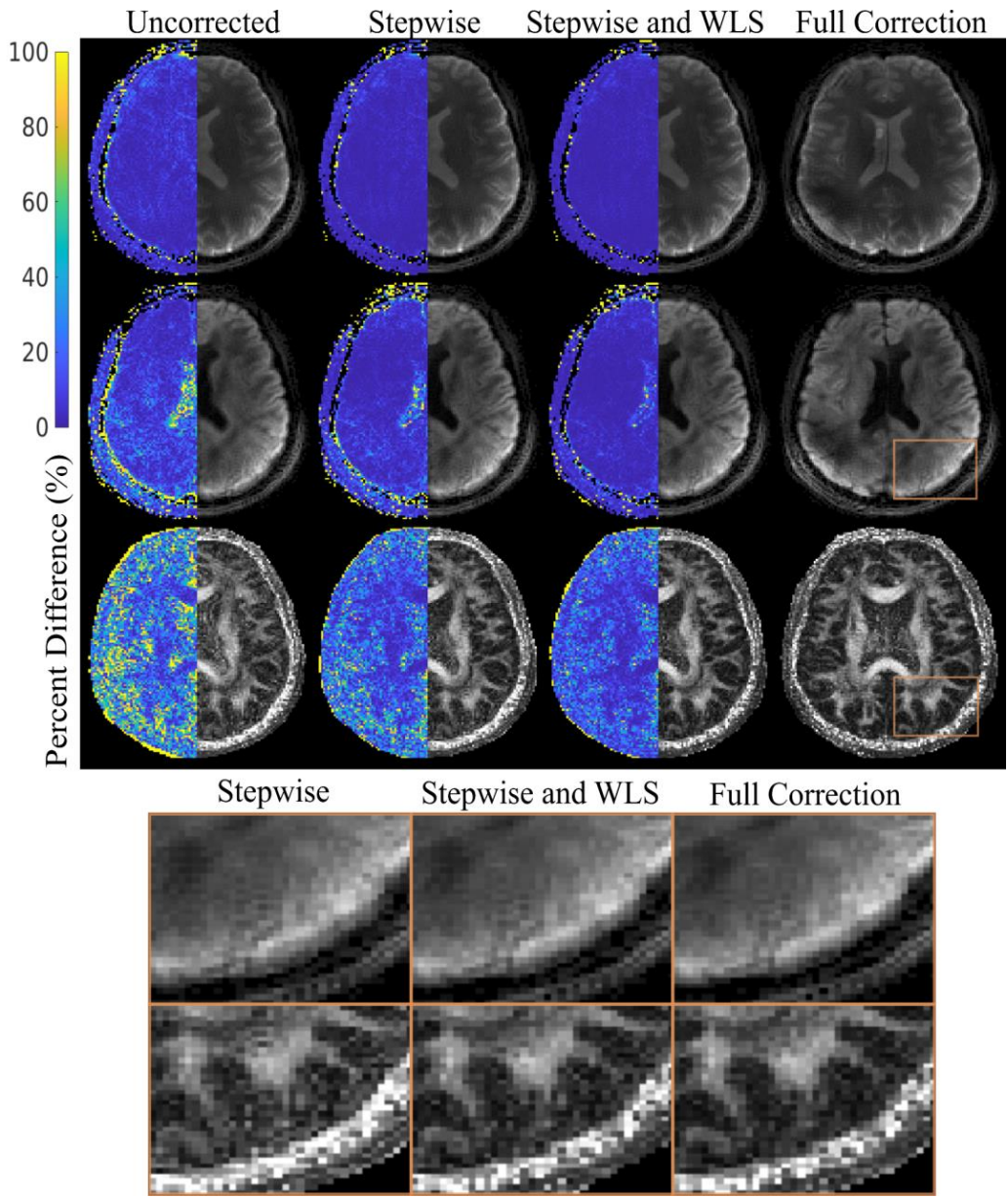

**Figure 6** Non-diffusion weighted images (b = 0 s/mm$^2$, top), single-direction DWI (b = 1000 s/mm$^2$, middle), and FA maps (bottom) reconstructed from single-shot spiral acquisitions (R = 2) using second order concurrent monitored field dynamics. Comparison of image quality for images incorporating uncorrected, stepwise-only corrected, stepwise + WLS, and fully corrected (stepwise + WLS + WPR) k-coefficients. Difference images calculated with respect to fully corrected images are shown in the left hemisphere of each image. Significant DWI blurring reduction and improved FA map integrity is seen when introducing stepwise correction. Zoom-ins further reveal more subtle blurring reduction and improved FA map quality when adding WLS correction, followed by WPR correction. An animation showing this comparison more clearly is shown in Animation S3.

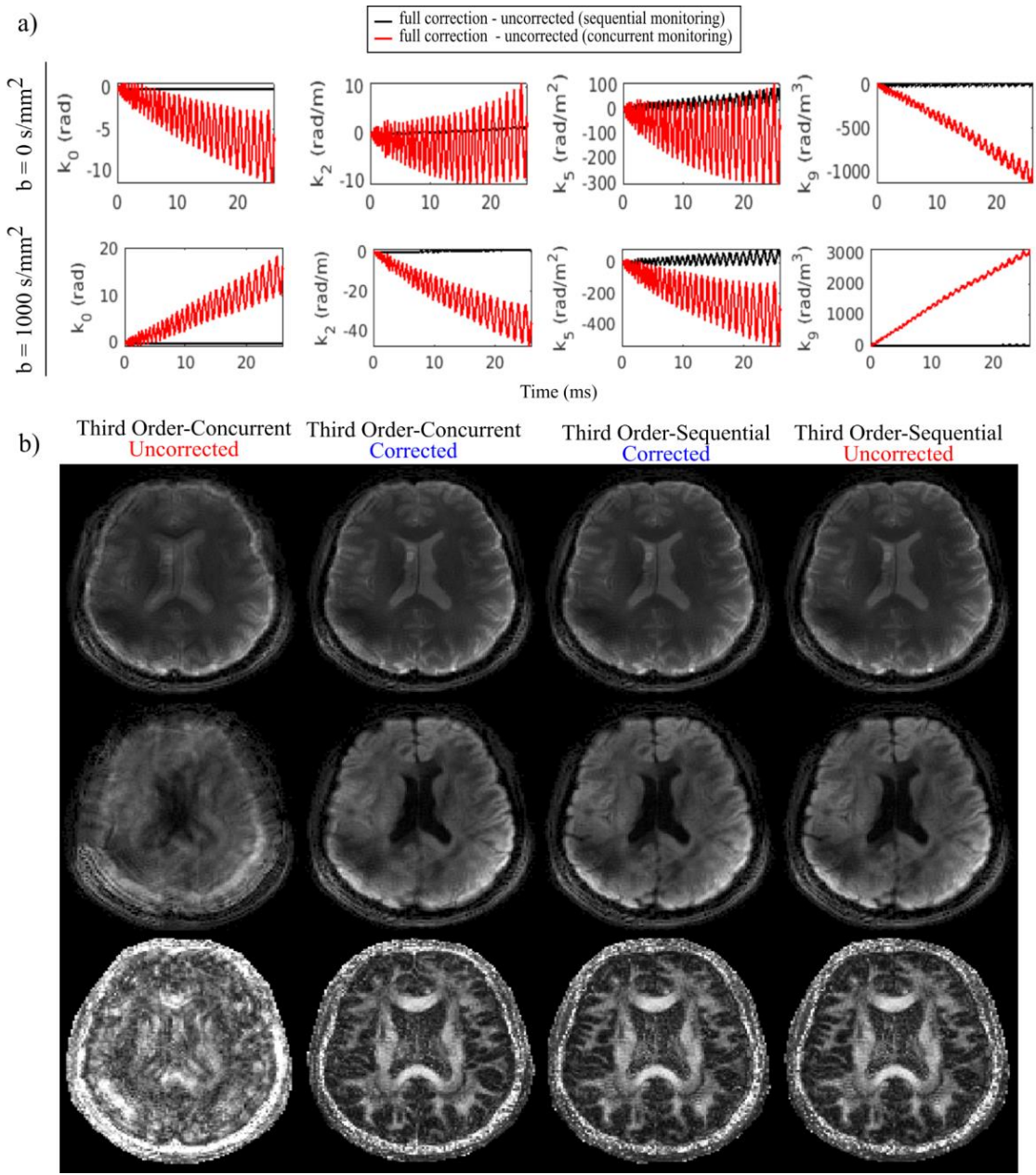

**Figure 7** Comparisons of concurrent and sequential field dynamic correction. (a) Difference in third-order fully corrected and uncorrected sequentially monitored (black) and concurrently monitored (red) k-coefficients (R = 2). (b) Non-diffusion weighted images (b = 0 s/mm$^2$, top), single-direction DWI (b = 1000 s/mm$^2$, middle), and FA maps (bottom) reconstructed using (left to right) concurrently monitored third order uncorrected k-coefficients, concurrently monitored third order corrected k-coefficients, sequentially monitored third order corrected k-coefficients, and sequentially monitored third order uncorrected k-coefficients. While large reconstruction errors are introduced by corrupted third order concurrently

monitored field dynamics, sequentially monitored field dynamics do not exhibit any observable reconstruction errors, given the optimized fitting scenario prior to correction.

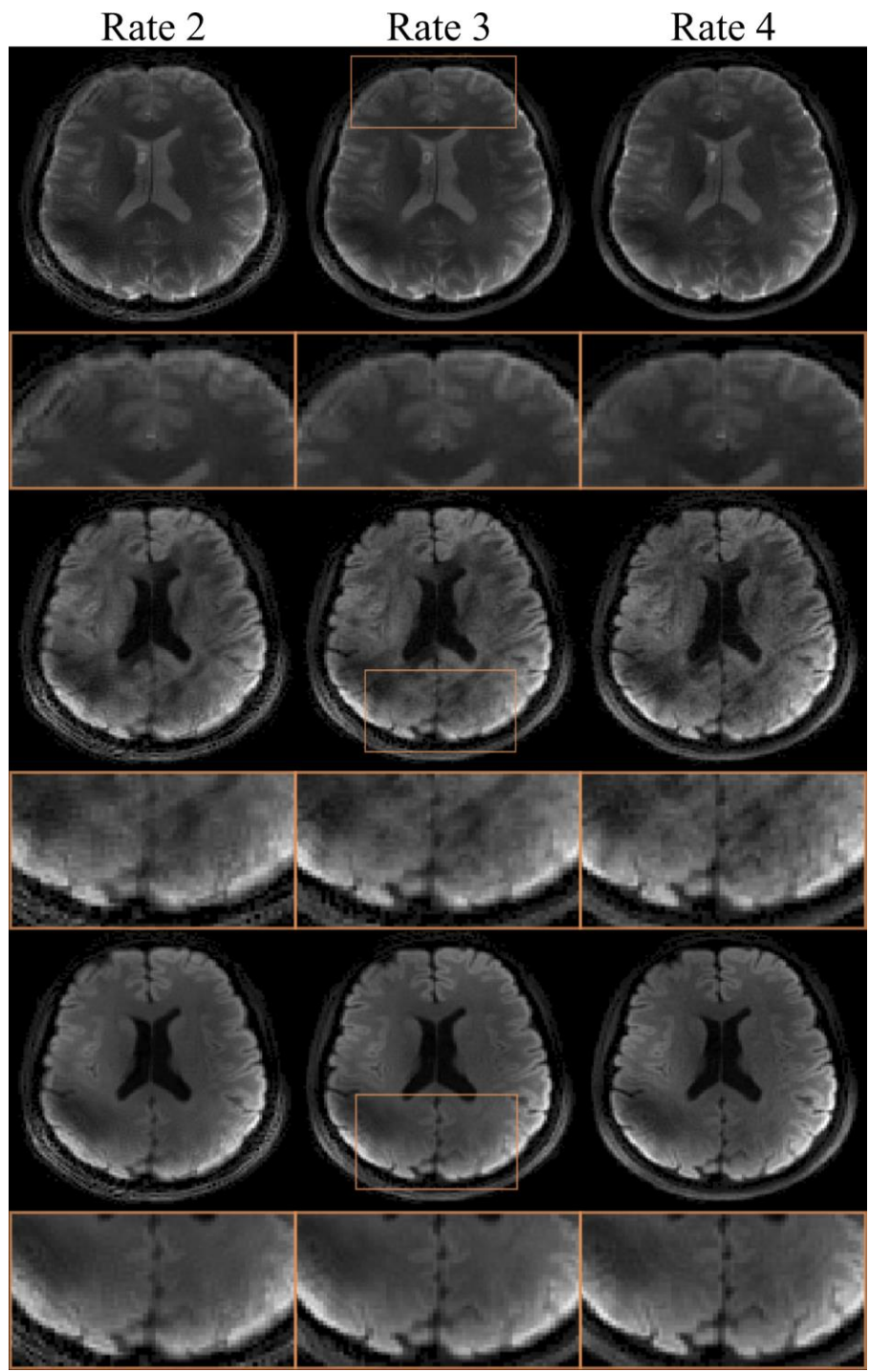

**Figure 8** Comparison of non-diffusion weighted images, single-direction DWI and average DWI reconstructed from single-shot spiral acquisitions using acceleration factors of 2,3 and 4 (left to right

respectively). Fully corrected second order concurrent monitored field dynamics were used in the expanded encoding model. Zoom-ins shown under each respective image. Reduction in blurring can be seen as the acceleration rate is increased, which is likely due to decreased $T_2^*$ blurring effects at shorter readout times, and a reduced sensitivity to remaining k-coefficient errors that may not be completely resolved by the proposed correction algorithm.

# Supporting Information

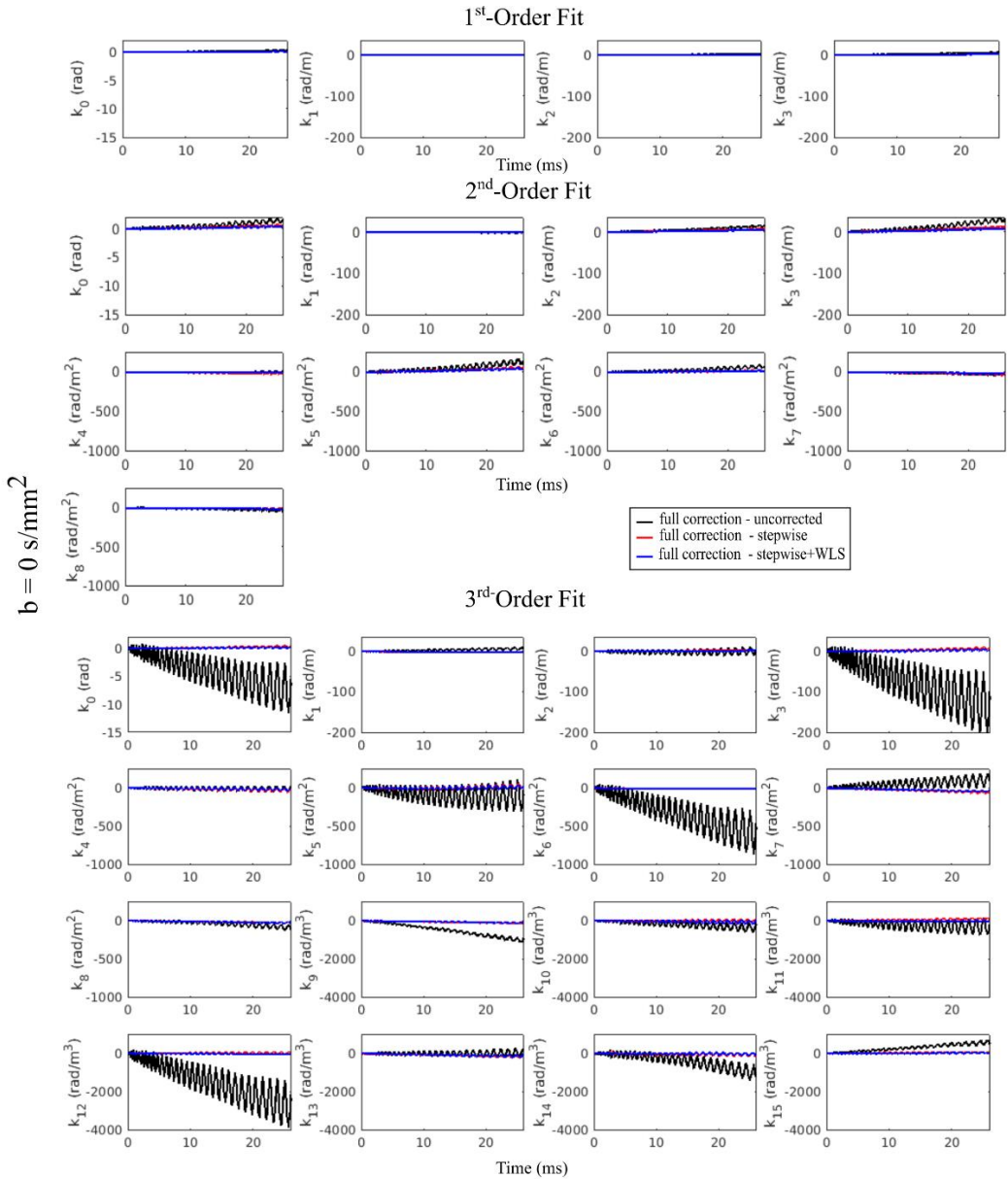

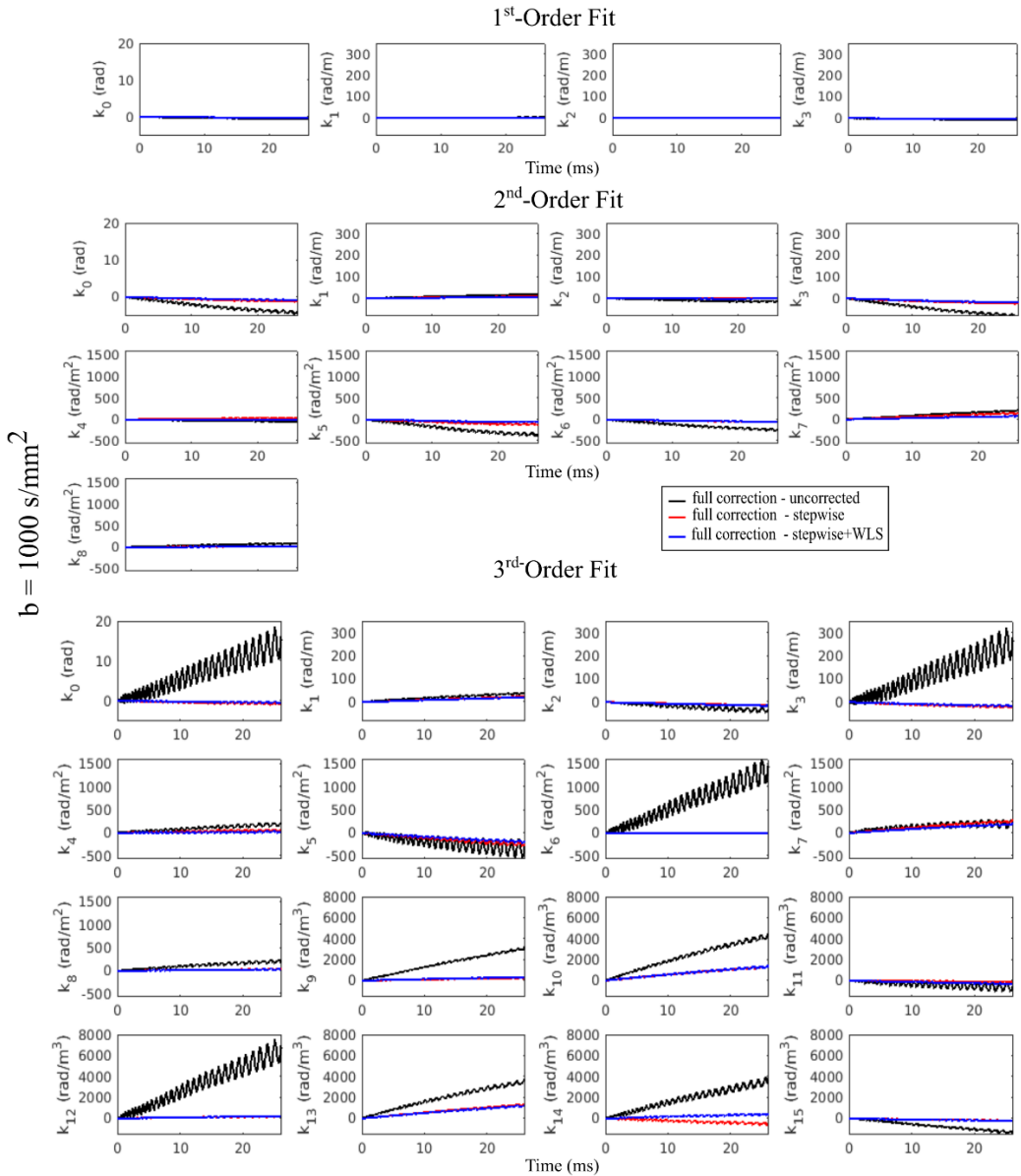

**Figure S1 (a) - Difference Profiles All Coefficients** Difference between the reference fully corrected k-coefficients and uncorrected coefficients (black), coefficients corrected using only stepwise correction (red), and coefficients corrected using stepwise + WLS correction (blue), for first through third order fits. Only black and blue curves displayed for the first order fit since there is no stepwise correction to be performed. First figure monitored a $b_0$ single-shot spiral acquisition, while the second figure monitored a diffusion weighted (b = 1000 s/mm$^2$) single-shot spiral acquisition (R = 2). The $0^{th}$-$2^{nd}$ order uncorrected differences increase as the fitting order is increased as a result of more complex fitting when higher orders are introduced. Moreover, the differences in uncorrected and corrected k-coefficients are generally greater for the diffusion weighted acquisitions because the diffusion weighted acquisitions likely exhibit more phase errors from stronger diffusion eddy currents. Larger differences are typically observed for terms that exhibit z-dependency.

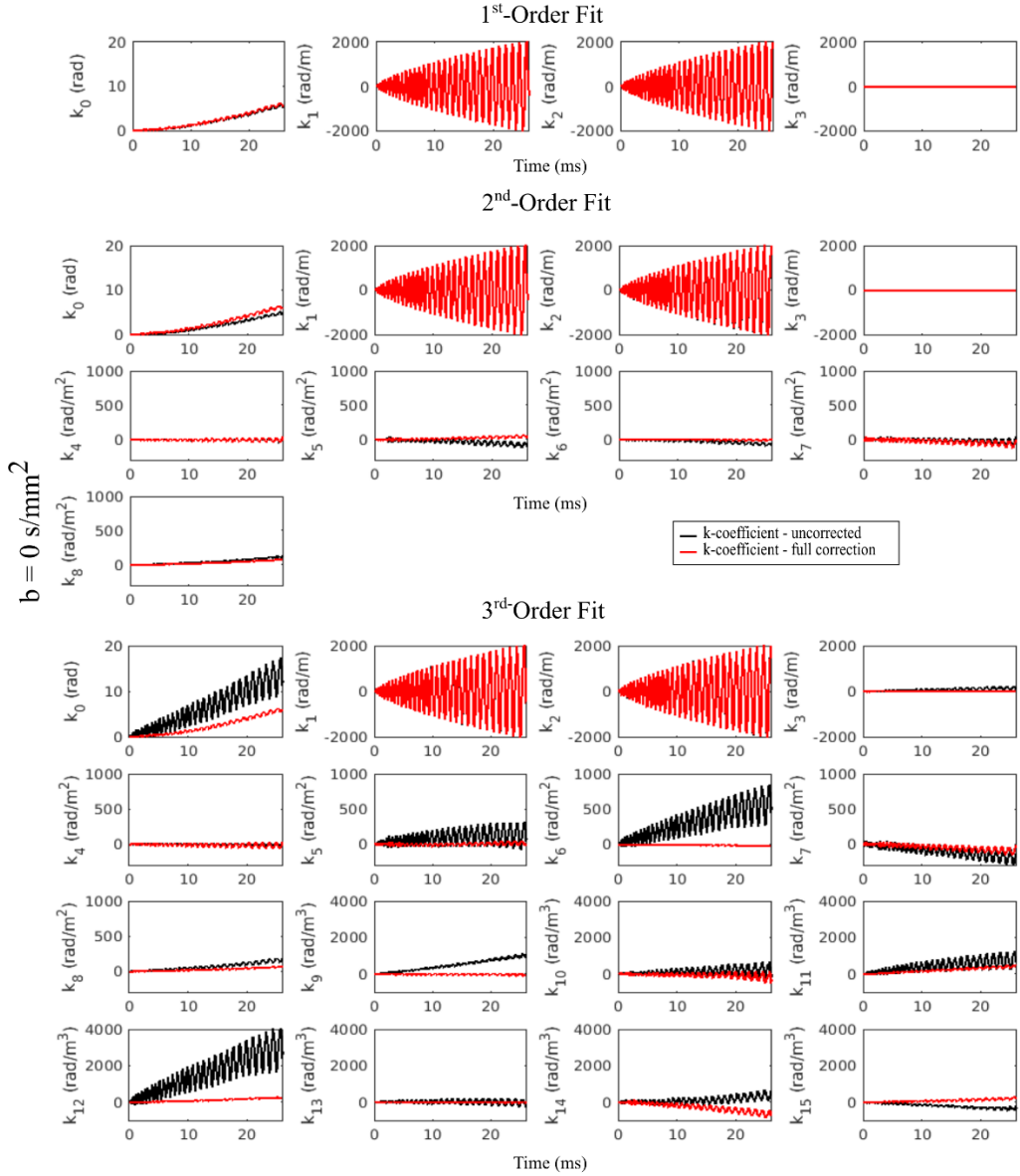

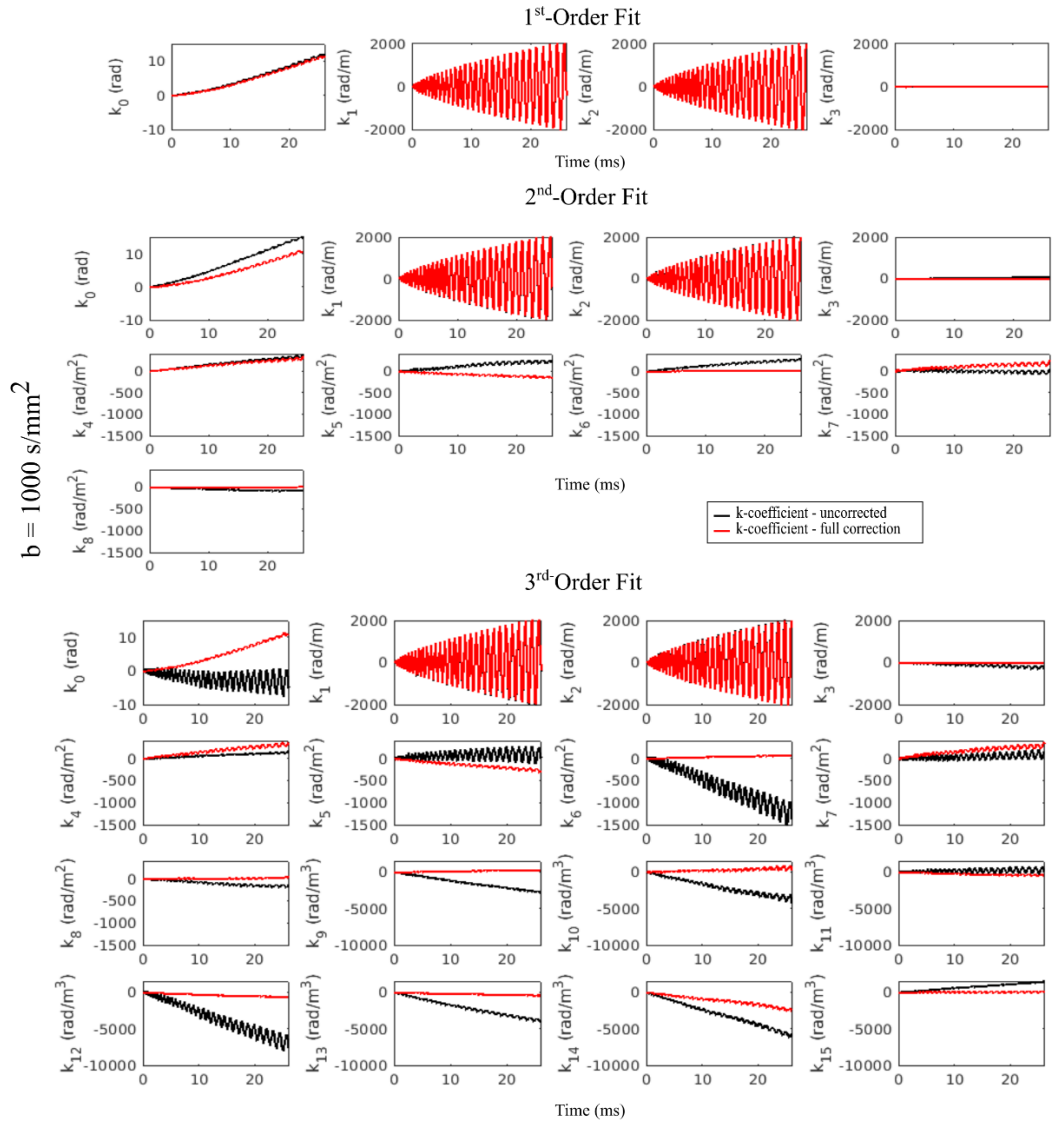

**Figure S1 (b) - k-coefficient Profiles All Coefficients** Uncorrected k-coefficients (black) and fully corrected k-coefficients (red), for first through third order fits. First figure monitored a $b_0$ single-shot spiral acquisition, while the second figure monitored a diffusion weighted (b = 1000 s/mm$^2$) single-shot spiral acquisition (R = 2). Larger phase accrual typically observed for uncorrected higher order terms, which are reduced with correction.

**Animation S2 - Order Comparison** Video comparison of first and second order corrections. Improved DWI and FA map quality are observed when using corrected first order field dynamics compared to uncorrected first order. This is further improved when using corrected second order fits.

**Animation S3 - Correction Technique Comparison** Video comparison of DWI and FA map quality when using no correction, subsets, and full use of the correction algorithm when performing second order fits of concurrently monitored field dynamics. Stepwise correction provides the most correction benefit, followed by incremental improvement with each successive correction.

**Animation S4 - EPI Results** Video includes difference profiles in second-order fully corrected k-coefficients and uncorrected k-coefficients (black), coefficients corrected using only stepwise correction (red), and coefficients corrected using stepwise + WLS correction (blue), for concurrently monitored k-coefficients from non-diffusion weighted (top row) and diffusion weighted (bottom row) EPI acquisitions (R = 2). Non-diffusion weighted images (b = 0 s/mm$^2$, left), single direction DWI (b = 1000 s/mm$^2$, middle), and FA maps (right) reconstructed from EPI acquisitions using concurrently monitored third through first order uncorrected and fully corrected k-coefficients, respectively. Fitting errors in EPI are shown to primarily translate as bulk image shifts in the phase encoding direction, which lead to erroneous FA map quality.

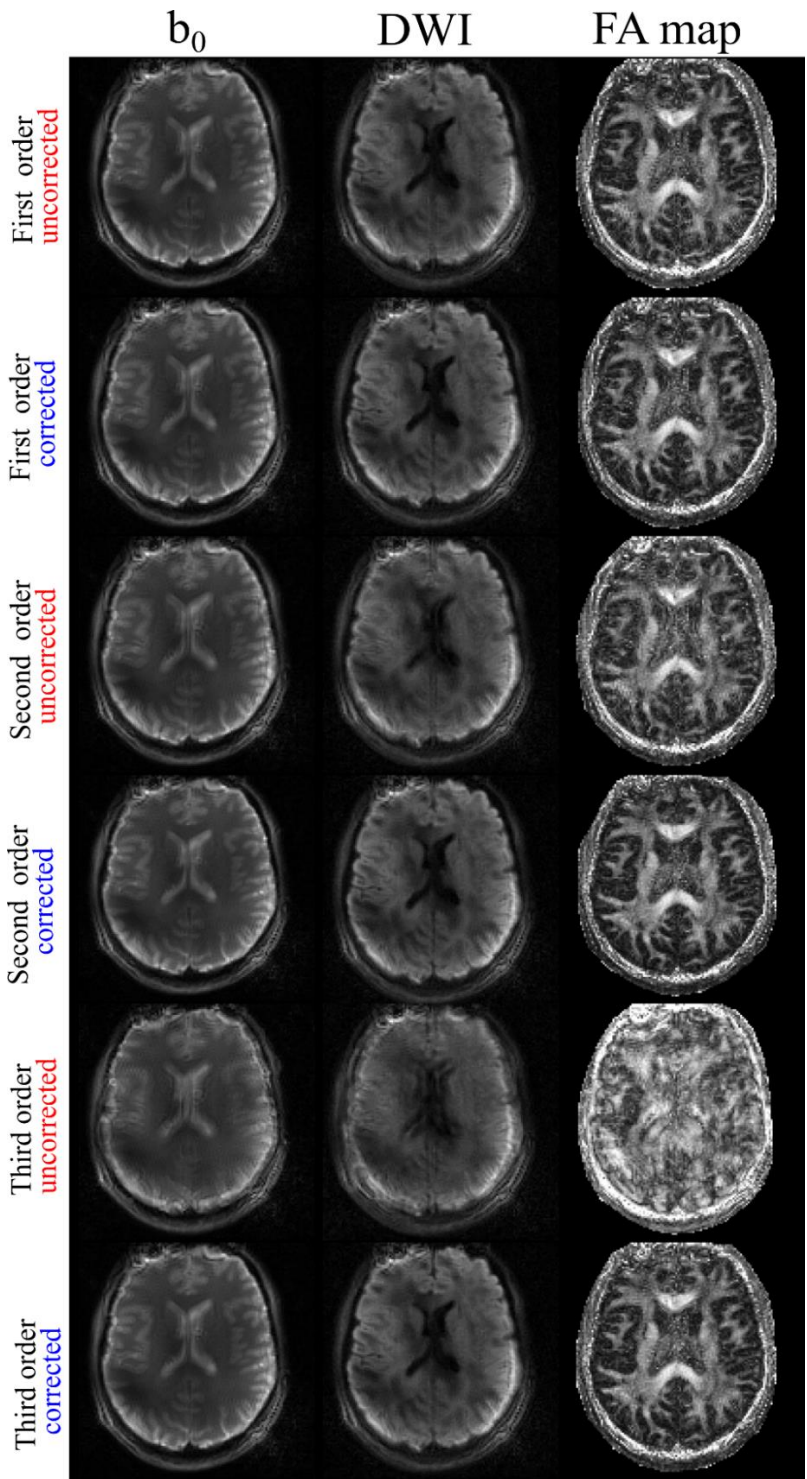

**Figure S5 – Algorithm Correction Performance: Volunteer 2** Image reconstructions from a second volunteer. Non-diffusion weighted images (b = 0 s/mm², left), single-direction diffusion weighted images (DWI) (b = 1000 s/mm², middle), and fractional anisotropy (FA) maps (right) reconstructed from single-shot spiral acquisitions (R = 2) using first through third order concurrent monitored field dynamics.

Reconstructions incorporated uncorrected k-coefficients and corrected k-coefficients using the proposed stepwise, WLS, and WPR correction techniques.

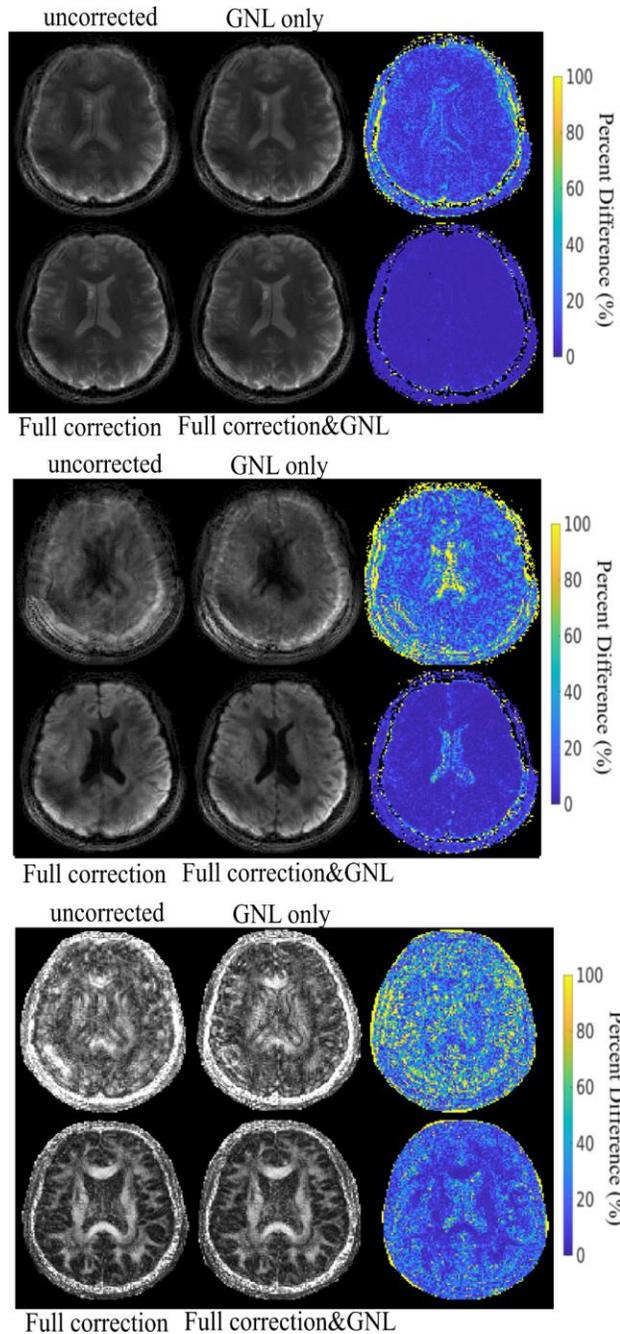

**Figure S6 - Explicit Gradient Nonlinearity Correction** Non-diffusion weighted images (b = 0 s/mm$^2$, top), single-direction DWI (b = 1000 s/mm$^2$, middle), and FA maps (bottom) reconstructed from single-shot spiral acquisitions using third order concurrent monitored field dynamics. Shown is a comparison of image quality for images incorporating uncorrected (upper left quadrant), gradient nonlinearity (GNL) corrected (upper right quadrant), fully corrected: stepwise, WLS, and WPR correction (bottom left quadrant), and fully corrected plus GNL correction (bottom right quadrant) k-coefficients. Difference

images calculated between images in each row. By itself, GNL correction improves image quality, but not sufficiently. Applying GNL correction in addition to the conventional three-component correction algorithm presents negligible difference in image quality.